\documentclass[12pt]{iopart}
\input epsf

\def\msol{M_\odot}
\def\mj{M_J}

\def\te{T_{\rm eff}}
\def\gcc{\rm g.cm^{-3}}
\def\simgr{\,\hbox{\hbox{$ > $}\kern -0.8em \lower 1.0ex\hbox{$\sim$}}\,}
\def\simle{\,\hbox{\hbox{$ < $}\kern -0.8em \lower 1.0ex\hbox{$\sim$}}\,}

\begin{document}

\title[The physics of brown dwarfs]{The Physics of Brown dwarfs}
\author{ G. Chabrier}
\address{Centre de Recherche Astrophysique de Lyon (UMR CNRS 5574),\\ 
 Ecole Normale Sup\'erieure de Lyon, 69364 Lyon
Cedex 07, France\\}

\begin{abstract}
We briefly outline the physics underlying the mechanical and thermal properties of brown dwarfs, which characterizes their interiors and their atmospheres.
We mention the most recent improvements realyzed in the
theory of brown dwarfs and the connection with experimental and observational
tests of this theory.
\end{abstract}

\section{Introduction}

A general outline of  the basic physics entering the structure and the evolution
of brown dwarfs (BD) can be found in the previous reviews of Stevenson (1991)
and Burrows \& Liebert (1993). Important innovations have occured in the field since
then, one of the least negligible being the {\it discovery} of bona-fide brown dwarfs
(Rebolo et al., 1995; Oppenheimer et al., 1995). An increasing number of these
objects have now been discovered either as companions of stars, as members of
young clusters or as free
floating objects in the Galactic field (Ruiz et al., 1997; Delfosse et al., 1997). On the other hand the theory has
improved substantially within the past few years and can now be confronted
directly to observations and even to laboratory experiments, as will be
shown below. It is thus important to reconsider the previous reviews in the
light of these observational and theoretical progress and to update our
knowledge of the structure and the evolution of BDs. This is the aim of
the present review.

BDs are objects not massive enough to sustain hydrogen burning in their
core and thus to reach thermal equilibrium, defined as $L=L_{nuc}$ where
$L_{nuc}=\int_0^M \epsilon dm$ is the nuclear luminosity and $\epsilon$ is
the nuclear reaction rate per unit mass. This hydrogen burning minimum mass $M_{HBMM}$ depends on
the internal composition of the object, in particular the abundance (by mass) of
helium (Y) and heavier elements (Z). For abundances characteristic of the solar
composition, typical of the Galactic disk population, $Y_\odot$=0.27 and $Z_\odot$=0.02, this minimum mass is
$M_{HBMM}\sim 0.072\, \msol$, whereas for compositions characteristic of the
Galactic halo (Y=0.25; $Z\sim 10^{-2}\times Z_\odot$) $M_{HBMM}\sim 0.083\,\msol$ (Chabrier \& Baraffe, 1997; Baraffe et al., 1997) and for the zero-metallicity limit ($Z=0$), $M_{HBMM}\sim 0.09\, \msol$ (Saumon \etal., 1994).

The minimum mass for BDs is presently  undetermined and masses as small as a
Jupiter mass ($10^{-3}\msol$) are not excluded in principle. The divided line
between BDs and giant planets (GPs) is still unclear and stems essentially
from their formation processes : hydrodynamic collapse of an interstellar
molecular cloud for BDs, like for stars, accretion of heavy elements
in a protostellar disk for the formation of planetesimals which eventually become
dense enough to capture hydrogen and helium and form gazeous planets. The
accretion scenario rather than  the collapse scenario for the formation of
planets is supported by the distinctly supersolar average abundance of heavy elements in Jupiter and Saturn, although there is only indirect 
evidence for the presence of the central rocky core through the modeling of
the centrifugal moments. The
border line between these two scenarios is presently unknown and involves most likely
complex dynamical and non-linear effects. Extra-solar planets with
masses as large as $\sim 40\,\mj$ and BDs with masses as low as
$\sim 30\,\mj$ have now been discovered. Except for this formation process and for the presence
of a central rock/ice core, the physics and the observational signature of BDs and
GPs is very similar. Since a complete review is devoted to GPs (Stevenson, this issue), the
present one will be devoted to BDs.
In the present survey, I will focus on the most recent improvements realyzed in
the physics of the interior and the atmosphere of BDs. I
will also mention the physics underlying the so-called Lithium-test, which provides a
powerful independent determination of the substellar nature and the age of a
putative BD. The aim of the present review is not to present detailed calculations
(which can be found in the various mentioned references) but rather to catch the
underlying physics entering the structure and the evolution of BDs.

\section {Interior of brown dwarfs.
The hydrogen equation of state}

Central conditions for massive BDs are typically $T_c\simle 10^5$ K and
$\rho_c \sim 10^2$-$10^3$ $\gcc$. Under these conditions, the average ion
electrostatic energy $(Ze^2)/a$, where $a=({3\over 4\pi}{V\over N_i})^{1/3}$
is the mean interionic distance, is several times the average kinetic energy
$kT$, characterizing a strongly coupled ionic plasma with a coupling parameter $\Gamma_i=(Ze)^2/akT > 1$.
The temperature is of the order of the electron Fermi temperature $kT_F$ and
the average inter-electronic distance a$_e$ is of the order of both the Bohr
radius $a_e\sim a_0$ and the Thomas-Fermi screening length $a_e\sim a_{TF}$. We thus
have  to deal with a partially degenerate, strongly correlated, polarizable
electron fluid. The temperature in the envelope is $kT \simle $1 Ryd, so we
expect electronic and atomic recombination to take place. At last the electron average binding energy is or the order of
the Fermi energy $Ze^2/a_0 \sim \epsilon_F$ so that {\it pressure}-ionization
is taking place along the internal density profile.

\begin{figure}
\begin{center}
\epsfxsize=95mm
\epsfysize=95mm
\epsfbox{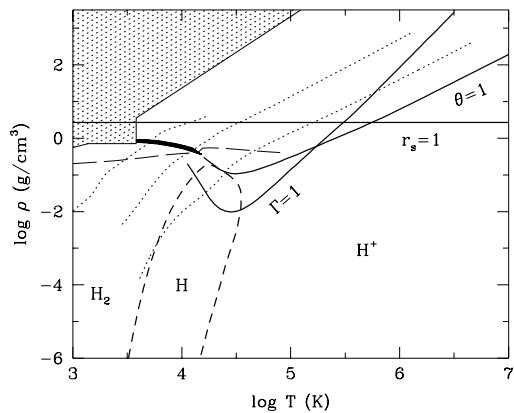}
\end{center}
\caption{Phase diagram of hydrogen in $T-\rho$.  The coexistence curve of the plasma phase transition
(PPT) appears at the left of center as a black solid line which ends at the critical point.  
Curves of constant ionic
($\Gamma$) and electronic ($r_s$) plasma coupling parameter and
 electron degeneracy parameter $\theta=T/T_F$
are
shown.
Regions dominated by molecules, 
atoms and ionized H are labeled  and delimited by a curve which corresponds to
50\% dissociation and ionization.  The calculated shock
Hugoniot corresponding to the experiments is shown by the long-dash curve
across the diagram.
Finally, the thin dotted lines show the
internal structure profiles of several astrophysical bodies (from left
to right), Jupiter, a 7 Ga brown dwarf of $0.055\,M_\odot$ and a $0.1\,\msol$ star.
The EOS model is invalid in the hashed region.}
\end{figure}

Recently laser-driven shock-wave experiments have been conducted at Livermore
(Da Silva et al., 1997; Collins \etal, 1998) which probe directly the thermodynamic properties of
dense hydrogen under conditions
characteristic of BDs and GPs. The relevance of the Livermore experiments for
the interior of these objects can be grasped from Figure 1.
About a decade ago, Saumon \& Chabrier (1991, 1992; SC) have developed a free energy model aimed
at describing the thermodynamic properties of strongly interacting
H$_2$ molecules, H atoms, H$^+$ protons and electrons under such
astrophysical conditions. The abundances of each species derive from the free energy minimization :

\begin{eqnarray}
\delta F(N_{H_2},N_H,N_{H^+},N_{e^-},V,T)=\Sigma_i{\partial F\over \partial N_i} \delta N_i=0
\end{eqnarray}

This model relies on the so-called chemical picture, which assumes that the species remain distinct even at high-density. This requires the
knowledge of the interparticle
potentials $\phi_{H_2H_2}, \phi_{H_2H}$ and $\phi_{HH}$. Since N-body effects
strongly modify the interaction between particles at high-density, "effective"
pair-potentials can be derived from the experimental hugoniots. These effective
pair-potentials mimic the softening of the interaction due to the surrounding
particles and thus retain some density-dependence in the characteristic interactions between
the main species. At the time the SC formalism was derived, the only available
shock-wave experiments were the ones by Nellis et al. (1983). Since molecular
dissociation was negligible under these conditions, only an effective potential
between H$_2$ molecules could be derived. The H$_2$-H and H-H potentials were
taken as ab-initio potentials. The more recent experiments by Weir et al. (1996) reach higher pressures and substantial molecular dissociation is
inferred from these experiments ($X_H > 20\%$). This allows us to derive
effective potentials for $\phi_{H_2H_2}$ and $ \phi_{H_2H}$ as well and thus to
update the SC model.

The recent laser-driven experiments have shown that the agreement between the {\it predictions}
of the SC model and the data is excellent. In particular the strong compression
factor arising from hydrogen pressure-dissociation and ionization
observed in the experiment ($\rho/\rho_i \sim 5.8$) agrees well with the predicted theoretical value (Saumon et al., 1998). The
compression is slightly underestimated in the theory and starts at a slightly
too large pressure. This reflects the underestimated degree of dissociation in
the model, which stems from the too repulsive (ab-initio) $\phi_{H_2H}$ potential
at the time the SC model was elaborated. This shortcomings is resolved when including the afore-mentioned new effective H$_2$-H$_2$ and H$_2$-H
potentials. Eventually full ionization is reached
at very high pressure ($P\sim 10$ Mbar), characterized by the asymptotic
compression factor $\rho_f/\rho_i =4$ for a monoatomic fully dissociated proton
fluid. 
These results show that, although this "chemical" model certainly does not pretend
to give an exact, complete description of all the interactions in the high-pressure strongly correlated fluid, it very likely retains the main physics underlying the phenomenon of pressure-dissociation/ionization.

It is worthnoting that a similar strong compression factor is obtained also with
the so-called fugacity expansion scheme, in principle exact in the strongly
dissociated regime (Rogers and Young 1997), although this scheme fails at lower
density when substantial recombination occurs.

One of the most striking features of the SC theory is the prediction of a first-order so-called plasma phase transition (PPT) between a molecular state and a
plasma state for the pressure-ionization of hydrogen, similar to the
one predicted originally by Wigner \& Huntington (1935). However it is important to
stress that the PPT in the SC model arises from first-principle thermodynamical instability of
the one single free energy model ($(\partial P/\partial \rho)_T <0$) and not from the comparison between two
different free energy models. The new SC EOS, incorporating the new potentials,
still predicts a PPT, although with a critical point slightly cooler than
predicted previously, namely $T_c=14600$ K, $P_c=0.73$ Mbar (Saumon et al., 1998).
In order to really nail down the existence of the  PPT, we have calculated
 a second-shock Hugoniot reflected from the principal one, which should be
realizable in a near future (Saumon et al., 1998).
Such an experiment should confirm or rule out definitely the presence of the PPT.

 The main question about the PPT is : if it exists, what is its nature ? This question
 has been addressed to some extend in Saumon \& Chabrier (1992). If the PPT
 exists, it stems very likely from the large difference between a molecular
state characterized by a
strongly repulsive potential and a plasma state characterized by a soft Yukawa-like potential. Given the large difference between these two potentials,
 and thus the respective available phase spaces, we can expect a discontinuity in the interaction
 energy and thus an abrupt change in the two-particle distribution function.
 This behaviour is observed in recent path-integral Monte-Carlo simulations
 (Magro et al., 1996; Ceperley, this issue). In terms of ground state energies,
 this translates into the large energy barrier between the ground state energy
 of an H$_2$-like system (H$_2$ or H$_2^+$) and an H$^+$-like system. In terms of correlation
 lengths that characterize the many-body effects, the system will collapse from
 a dense molecular phase characterized by a length $\lambda_{H_2}\sim$ a few $a_0$ into a plasma phase characterized by a length $\lambda_{H+} << \lambda_{H_2}$. The underlying critical quantity will be the electron correlation length,
with a critical percolation from a "bound-electron"-like value to a "free-electron"-like value.
In this sense the PPT resembles the metal-insulator transition in metals
associated with the liquid-vapor transition (Hensel, this issue), leading
eventually to a polarization catastrophe (Goldstein \& Ashcroft, 1985). The
effect is likely to be more dramatic for hydrogen because of the absence of
core electrons.

In this sense, the conductivity measurements of dense fluid hydrogen by Weir et al. (1996) do not rule out the PPT. The conductivity 
exhibits a plateau with $\sigma \sim 2000\,(\Omega.cm)^{-1}$ up to the
highest pressure reached, $P\sim 1.8$ Mbar. This is still orders of magnitude
smaller than  the conductivity characteristic of a fully dissociated plasma
phase, $\sigma \sim 10^5\,(\Omega.cm)^{-1}$ (Stevenson \&
Ashcroft, 1974) and is consistent with conduction being due to delocalized
electrons from $H_2^+$. This does not preclude a
{\it structural} transition like the PPT at higher pressures.

If the PPT exists it can have important consequences for BDs and GPs. The interior of
these objects is essentially isentropic. Since the signature  of a first-order
transition is a density and entropy discontinuity, integration along the internal adiabat from
the observed outer conditions yields different central conditions with and
without PPT (Chabrier et al., 1992). In principle the signature of the PPT in the interior of
GPs like Jupiter and Saturn could be observed from the analysis of p-mode
oscillations (Marley, 1994; Gudkova et al., 1995). However this requires very accurate observations
of high-degree modes, a difficult observational task.
The PPT bears also important consequences on the evolution of these objects.
Since by definition BDs and GPs do not sustain hydrogen burning, application
of the first and second principles of thermodynamics yields the following equation for their evolution :

\begin{eqnarray}
L=-{d\over dt}\int_0^M (\tilde u+{P\over \rho^2}{d\rho \over dt})dm=
-\int_0^M T{d\tilde s\over dt} dm 
\end{eqnarray}

where $L$ is the luminosity, $\tilde u$ and $\tilde s$ the  specific internal energy
and entropy, respectively. 
If the PPT exists, an additional term, namely the latent heat of the phase
transition, must be added to the previous equation :

\begin{eqnarray}
L^\prime = L + \int_{\Delta m} T {d\Delta \tilde S\over  dt} dm
\end{eqnarray}

This effect was first pointed out by Stevenson \& Salpeter (1977) and
examined in detail by Saumon et al. (1992). 

\section{The atmosphere of brown dwarfs}

\subsection{Spectral distribution}

The photosphere is defined as the location where the photon
mean free path is of the order of the mean interparticle distance,
i.e. $l_\nu \sim 1/(\bar \kappa \rho)\sim a \propto \rho^{-1/3}$,
where $\bar \kappa \sim 1 cm^2/g$ is the  mean absorption coefficient (opacity). This equality yields $l_\nu \sim a \sim 1$ cm.
In terms of the dimensionless optical depth $\tau = z/l_\nu$, where $z$
is the depth of the atmosphere, equilibrium between internal
and gravitational pressure yieds :

\begin{eqnarray}
d\tau=- (\rho \bar \kappa)\, dz= \bar \kappa {dP\over g},
\end{eqnarray}

where $g=GM/R^2$ is the surface gravity. For BDs, $M\simle 0.1\,\msol$,
$R\sim 0.1\,R_\odot$, $g\simle 10\times g_\odot$.
This yields $P_{ph}\sim g/\bar \kappa \sim 10$ bar at the
photosphere, and $\rho_{ph}\sim 10^{-5}-10^{-4}\,\gcc$.
Collision effects are significant under these conditions. Therefore
thermodynamic equilibrium can be safely assumed near the photosphere.
The bad news is that collision effects can induce dipoles between
molecules, e.g. H$_2$ or He-H$_2$, which otherwise would have only
quadrupolar transitions. This so-called collision-induced absorption
(CIA) between roto-vibrational states ($v\rightarrow v^\prime$) of e.g. 2
H$_2$ molecules (1 and 2) can be written in terms of the 2-body absorption 
(see e.g. Borysow et al., 1985):

\begin{eqnarray}
\kappa_{H_2H_2}&= \Sigma_{v_1,v^\prime_1} \Sigma_{v_2,v_2^\prime}
\alpha_{H_2H_2}^{v_1,v_1^\prime,v_2,v_2^\prime} (\omega,T)\nonumber\\
&= n_{H_2}^2 {2\pi^2 \over 3\hbar  c} w(1-e^{-\hbar \omega / kT})
\Sigma_{v_1,v^\prime_1} \Sigma_{v_2,v_2^\prime}
g^{v_1,v_1^\prime,v_2,v_2^\prime}(\omega, T)
\end{eqnarray}

where $\omega=2\pi \nu$ is the angular frequency, $n_{H_2}$ is the
number density of hydrogen molecules and $g^{v_1,v_1^\prime,v_2,v_2^\prime}(\omega,T)$  is the spectral function.
We note the dependence on the square of the number abundance. As temperature
decreases below $\sim 4000$ K, an increasing number of hydrogen
molecules form and thus H$_2$ CIA-absorption becomes overwhelmingly important, a feature shared with giant planet and white dwarf atmospheres.
Since the CIA absorption of H$_2$ under the conditions of interest for BDs and GPs
takes place around 2.2 $\mu$m, energy conservation leads to a redistribution of the emergent radiative flux toward shorter
wavelengths (Saumon et al., 1994; Baraffe et al., 1997).
\bigskip

The effective temperature is defined as the integral of the Eddington flux
over the frequency spectrum :

\begin{eqnarray}
\te^4 = \sigma^{-1} \int H_\nu d\nu,
\end{eqnarray}

where $\sigma = 5.67\times 10^{-5}$ erg.cm$^2$.K$^4$.s$^{-1}$is the Stefan-Boltzman constant. BDs are characterized by
effective temperatures $\te \simle 2000$ K. At these temperatures, numerous
molecules like e.g. H$_2$, H$_2$O, TiO, VO, ... are stable and are the major
absorbers of photons. These strongly frequency-dependent opacity sources
yield a strong departure from a black-body energy distribution (see e.g. Figure 5 of Allard et al. (1997)). An updated detailed review of the physics of the atmosphere of
low-mass
stars and BDs can be found in Allard et al. (1997).

Below $T\sim 1800$ K, carbon monoxide $CO$ is predicted to dissociate and to form methane,
$CH_4$, as observed in Jupiter. This prediction has been confirmed
by the discovery and the spectroscopic observation of Gliese229B. The
presence of methane in its spectrum assessed unambigously its sub-stellar nature.
Consistent synthetic spectra and evolutionary calculations done both by the  Lyon
group (Allard et al., 1996) and the Tucson group (Marley et al., 1996)
yielded the mass determination of the object between $\sim $20 and 50 $\mj$,
the undermination in the mass reflecting the undetermination in the age of the
system.
\bigskip

At last below $\sim 2000$ K complex compounds (grains, also called
"clouds" by planetologists) condensate in the atmosphere (see e.g. Lunine et al., 1986; Tsuji et al., 1996). These grains will affect
the atmosphere in different ways. They first modify the EOS itself and thus the
atmospheric temperature/density-profile, and they also strongly affect the
atmospheric opacity and thus the emergent radiation spectrum. At last they will
produce an increase of the temperature in the uppermost layers of the atmosphere, the so-called backwarming (or
greenhouse) effect, destroying otherwise stable polyatomic
species. The condensation of the grains in a BD  atmosphere is illustrated in
Figure 2. Spectroscopic observations of different BDs at various effective
temperatures show evidence for an even more complicated problem, namely grain
diffusion (settling) in the atmosphere.

\begin{figure}
\begin{center}
\epsfxsize=85mm
\epsfysize=85mm
\epsfbox{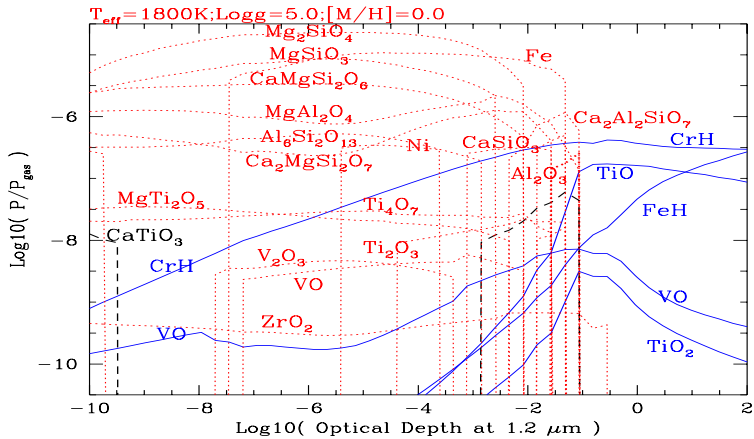}
\end{center}
\caption{
Relative abundances of gas phase 
(full lines) and crystallized species (dotted lines) across a T$_{\rm
eff}=1800$~K brown dwarf model atmosphere (after F. Allard).}
\end{figure}

\subsection{Energy transport}

The radiative transport equation reads :

\begin{eqnarray}
F_{rad}={4\over 3 \bar \kappa \rho} {d\over dr} (\sigma T^4)
\propto {\nabla T \over \bar \kappa}
\end{eqnarray}

\noindent for the radiative flux, while convective transport can be writen as :

\begin{eqnarray}
F_{conv}  \propto (\rho \, v_{conv})
\times (\tilde c_p \, \delta T)
\end{eqnarray}

where $v_{conv}$ is the convection velocity, typically a fraction of the speed of sound, $\tilde c_p$ is the matter specific heat
at constant pressure and $\delta T$ is the energy difference between the
convective eddy and the surrounding ambient medium. As the temperature
decreases below $\sim 5000$ K, which corresponds to a mass $m < 0.6\,\msol$,
H atoms recombine, $n_{H_2}$ increases, and so does $\bar \kappa$
through H$_2$ CIA-absorption (see above). The opacity increases by
several orders of magnitude over a factor 2 in temperature. On the other
hand, the presence of molecules increases the number of internal degrees
of freedom (vibration, rotation, electronic levels) and thus $c_p$.
These combined effects strongly favor the onset of convection in the
optically-thin ($\tau < 1$) atmospheric layers. This can be shown easily
from a stability (Schwarzchild) criterion analysis. Flux conservation thus
reads :

\begin{eqnarray}
\nabla (F_{rad} + F_{conv}) = 0
\end{eqnarray}

\noindent i.e. no radiative equilibrium.
The evolution of low-mass objects (low-mass stars, BDs, GPs) thus requires
to solve the complete set of the transfer equations and to use
consistent boundary conditions between the atmosphere and the interior
structure profiles (Chabrier \& Baraffe, 1997; Baraffe et al., 1995, 1997, 1998;
Burrows et al., 1997).

\section{Screening factors and the Lithium-test}

Since a BD, by definition, never reaches thermal equilibrium
($L\sim T\,dS/dt$), age is an extra degree of freedom, yielding an undetermination in the mass and/or age of an object for a given observed
luminosity and/or temperature. An independent
age-indicator is thus needed. The presence of Lithium
in the atmosphere of a cool object provides such an indication.
The signature of Lithium
absorption as a diagnostic for the sub-stellar nature of an
object was first pointed out by Rebolo
et al. (1992) while the measure of Lithium-depletion
as an age tracer was first used by Basri et al. (1996). 

The physics underlying the Lithium-test roots in dense plasma
physics and in the calculations of the so-called nuclear screening
factors for the nuclear reaction rate.

Primordial $^7Li$ is detroyed through the nuclear reaction
$^7Li + p \rightarrow 2^4He$. The reaction rate $R_0$ (in cm$^{-3}$$s^{-1}$) in the  vacuum
is given by the usual Gamow theory $R_0\propto e^{-3 \epsilon_0/kT}$
where $\epsilon_0$ corresponds to the Gamow-peak energy for non-resonant reactions, which
corresponds to the  maximum probability for the reaction.
However, as mentioned above, non-ideal effects dominate in the
interior of BDs and lead to polarization effects in the plasma.
These polarization effects due to the surrounding particles
yield an enhancement of the reaction rate, as first recognized by
Schatzman (1948) and Salpeter (1954).
The distribution of particles in the plasma reads :

\begin{eqnarray}
n(r) = \bar n e^{-Ze\phi(r)/kT}
\end{eqnarray}

\noindent with

\begin{eqnarray}
\phi(r) = {Ze\over r} \,+\, \psi(r)
\end{eqnarray}

\noindent where $\psi(r)$ is the induced mean field potential due
to the polarization of the surrounding particles. 
This induced potential lowers the Coulomb barrier between the
fusing particles and thus yields an {\it enhanced} rate in the
plasma $R=E\times R_0$ where

\begin{eqnarray}
E=lim_{r\rightarrow 0}\, \bigl\{ g_{12}(r) exp({Z_1Z_2e^2\over rkT})\bigr\}
\end{eqnarray}

\noindent is the enhancement (screening) factor and $g_{12}(r)$ the pair-distribution function.

\begin{figure}
\begin{center}
\epsfxsize=65mm
\epsfysize=65mm
\epsfbox{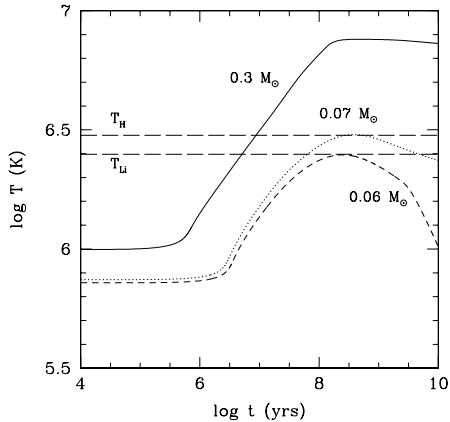}
\end{center}
\caption{Central temperature as a function of age for 3 different
masses respectively above, at the limit and below the hydrogen-burning
minimum mass. $T_H$ and $T_{Li}$ indicate the Hydrogen and Lithium burning
temperatures, respectively.}
\end{figure}

Under BD conditions, not only {\it ionic} screening must be included
but also {\it electron} screening, i.e. $E=E_i\times E_e$. Both effects
are of the same order ($E_i\sim E_e \sim$ a few) and must be included in the calculations for
a correct estimate of the Lithium-depletion factor
$[Li]_0/[Li]$, where $[Li]_0=10^{-9}$ denotes the primordial
Lithium-abundance.
This yields a Lithium-burning minimum mass $m_{Li}\sim 0.06\,\msol$
 (Chabrier \& Baraffe, 1997), {\it below}
the hydrogen-burning minimum mass, as illustrated in Figure 3. After the
common primordial deuterium burning phase, which lasts $\sim 10^6-10^7$ yr,
the central temperature evolves differently, depending on the mass of the
object. Note the strong age dependence of the Lithium-test : young {\it stars}
with an age $t\simle 10^8$ yr (depending on the mass) will exhibit Lithium,
whereas massive {\it brown dwarfs} within the mass range [0.06-0.07 $\msol$]
older than $\sim 10^8$ yr will have burned Lithium. The measure of Lithium depletion in the atmosphere
of low-mass objects, inferred from the width of the LiI line at 6708 $\AA$,
as an age indicator, is illustrated in Figure 4. This figure displays the evolution
of a 0.075 $\msol$ object, the H-burning limit for solar-abundances,
in the I-band magnitude. The left and right diagonal solid lines correspond to
50\% Li-depletion ($[Li]_0/[Li]=1/2$) and 99\% Li-depletion, respectively. Thus, for
say 120 Myr, the inferred age of the Pleiades cluster, objects brighter than
$M_I\sim 12.2$ will lie on the right-hand side
of the 99\%-depletion line and thus are predicted to show no Lithium in
their atmosphere and to be H-burning stars ($m\ge 0.075\,\msol$), whereas
objects fainter than this magnitude will all show {\it some} Lithium
and all be brown dwarfs ($m<0.075\,\msol$ for this age), with objects
fainter than $M_I\sim 12.6$ predicted to have retained more than half their
primordial Lithium-abundance. The horizontal lines show the observed
magnitudes of 4 different objects in the Pleiades, with available high-resolution spectra. All four confirm the
theory, with no Lithium observed for PL10, about 50\% depletion for
PL13 and negligible or no depletion for Roq13 and Teide1. Different isochrones for different masses can be superimposed on the same diagram and
analyzed similarly. This illustrates convincingly the powerful
diagnostic of Lithium as a mass and age indicator for low-mass stars and
brown dwarfs.

\begin{figure}
\begin{center}
\epsfxsize=65mm
\epsfysize=65mm
\epsfbox{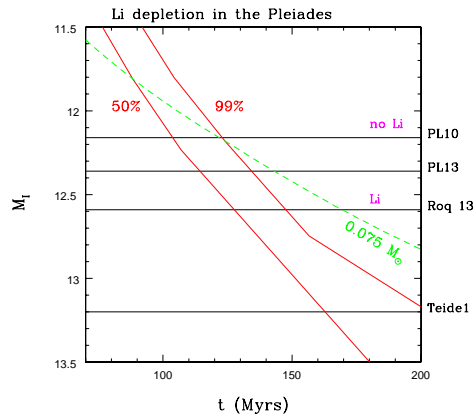}
\end{center}
\caption{Evolution of the absolute magnitude M$_I$ as a function of age.
The dashed line corresponds to the hydrogen-burning minimum mass, whereas
the diagonal solid lines correspond to the 50\% et 99\% Lithium-depletion
limit. The horizontal solid lines indicate the observed magnitudes of
different low-mass objects.}
\end{figure}

\section{Conclusion}

As a conclusion to this review, I will list a series of "homework
problems" related to BDs which illustrate the major problems to be addressed
in this field in a near future and which correspond to different domains
of physics or astronomy. This list is certainly not exhaustive.

\indent$\bullet$ {\it Dense matter physics}: As we have seen, BD interiors can now
be tested directly in laboratories and the EOS of these objects can be probed
by high-pressure experiments. More
experiments are needed in the complex regime of H-pressure dissociation/ionization with several unanswered questions. Does the PPT really exist ? Does it survive when 10\% helium
particles are present ? How does pressure-ionization of H affect the dynamo
process in BD and GP interiors ?
 
\indent$\bullet$ {\it Star formation process}:
Jeans stability analysis yields a minimum mass $m_{min}\sim
0.01\,\msol$, definitely in the BD domain (Silk, 1977). Is this mass the BD minimum mass ? Conversely what is the maximum mass
for planet formation ?
Does the Jeans criterion really apply for the formation of star-like objects ? What is the BD {\it mass function} in the Galaxy ?
 
\indent$\bullet$ {\it Evolution}:
The evolution of BDs is not hampered by any adjustable parameter, like for example in the treatment of convection for more massive stars which develop an
inner radiative core. The theory of BDs, and the comparison with  observation,
thus reflects the validity of the very physics entering the theory, both in
the atmosphere and in the interior. This theory can be tested directly now
by photometric and spectroscopic observations and must address new problems
like
e.g. the diffusion process of grains in the atmosphere or the magnetic field generation in active BDs. Conversely, the theory is now reliable
enough to provide useful
guidance for future observations.

\indent$\bullet$ {\it Galactic implication}:
The mass-to-light ratio  for BD, $(M/L)_{BD} 
 \simgr 10^4 (M/L)_\odot$, make BDs very promizing candidates to explain at least
the baryonic missing mass. Even though present estimates of their contribution
to the Galactic disk and halo mass seem to exclude this possibility
(Chabrier \& M\'era, 1997; M\'era et al., 1998), the
determination of their exact number- and mass-density in the Galaxy remains to
be determined accurately. Ongoing microlensing experiments sensitive to
hours and day event durations and ongoing wide field infrared projects
(e.g. DENIS, 2MASS) will certainly help nailing down this issue.

BDs thus present a wide variety of interest from basic physics to
Galactic implications and should remain a very active field.
\bigskip

\subsection{ Acknowledgments :} 
The results mentioned in this review to illustrate the physics of BDs arise from an ongoing collaboration with F. Allard, I. Baraffe and P.H. Hauschildt for the structure and the evolution of BDs, and with D. Saumon for the EOS. My profound
gratitude to these persons for their contribution to this review.

\section*{References}
\begin{harvard}
\item[] Allard, F., Alexander, D., Hauschildt, P.H. \& Starrfield, S., 1997, {\it ARA\&A} {\bf 35}, 137
\item[] Allard, F., Hauschildt, P.H., Baraffe, I. \& Chabrier, G., 1996,
 {\it Astroph. J.}, {\bf 424}, 333
\item[] Basri, G., Marcy, G., \& Graham, J. R., 1996,  {\it Astroph. J.}, {\bf 458}, 600
\item[] Baraffe, I., Chabrier, G., Allard, F. \& Hauschildt, P.H., 1995, {\it Astroph. J.}, {\bf 446}, L35
\item[] Baraffe, I., Chabrier, G., Allard, F. \& Hauschildt, P.H., 1997, , {\it Astron. \& Astroph.}, {\bf 327}, 1054
\item[] Baraffe, I., Chabrier, G., Allard, F. \& Hauschildt, P.H., 1998, , {\it Astron. \& Astroph.}, 337, 403
\item[] Borysow, A., Trafton, L., Frommhold, L. \& Birnbaum, G., 1985,  {\it Astroph. J.}, {\bf 296}, 644
\item[] Burrows, A.,
Marley, M., Hubbard, W.B., Lunine, J.I., Guillot, T.,
Saumon, D., Freedman, R., Sudarsky, D., Sharp, C. 1997,
{\it Astroph. J.}, {\bf 491}, 856
\item[] Burrows, A. \& Liebert, J., 1993, {\it Rev. Mod. Phys.}, {\bf 65}, 301
\item[] Chabrier, G \& Baraffe, I,, 1997, {\it Astron. \& Astroph.}, {\bf 327}, 1039
\item[] Chabrier, G \& M\'era, D., 1997, {\it Astron. \& Astroph.}, {\bf 328}, 83
 
\item[] Chabrier, G., Saumon, D., Hubbard, W.B. \& Lunine, J.I., 1992
 {\it Astroph. J.} {\bf 391}, 817
\item[] Collins, G.W., \etal, 1998, {\it Science}, {\bf 281}, 1178
\item[] Goldstein R.E. \& Ashcroft, N. W., 1985, {\it Phys. Rev. Lett.}, {\bf 55}, 2164
\item[] Da Silva, L.B., et al., 1997, {\it Phys. Rev. Lett.}, {\bf 78}, 483
\item[] Delfosse, X. et al., 1997,  {\it Astron. \& Astroph.}, {\bf 327}
\item[] Gudkova, T., Mosser, B., Provost, J., Chabrier, G., Gautier, D. \& Guillot, T., 1995,   {\it Astron. \& Astroph.}, {\bf 303}, 594
\item[] Lunine, J.I., Hubbard, W.B., and Marley, M.S., 1986, {\it Astroph. J.}, {\bf 310}, 238
\item[] Magro, W.R., Ceperley, D.M., Pierleoni, C., \& Bernu, 1996, {\it Phys. Rev. Lett.} {\bf 76}, 1240
\item[] Marley, M.S., Saumon, D., Guillot, T., Freedman, R., Hubbard, W.B., Lunine, J.I. \& Burrows, A., 1996, {\it Science}, {\bf 272}, 1919
\item[]  Marley, M.S., 1994, {\it Astroph. J.}, {\bf 427}, L63
\item[] M\'era, D., Chabrier, G. \& Schaeffer, R., 1998,  {\it Astron. \& Astroph.}, {\bf 330}, 937; , {\bf 330}, 953
\item[] Nellis, W.J., Mitchell, A.C., van Thiel, M., Devine, G.J., \& Trainor, R.J., 1983, {\it J. Chem. Phys.} {\bf 79}, 1480
\item[] Oppenheimer, B.R., Kulkarni, S.R., Nakajima, T. \& Matthews, K., 1995, {\it Science}, {\bf 270}, 1478
\item[] Rebolo, R., Mart\'in, E.L., Magazz\`u, A, 1992, {\it Astroph. J.}, {\bf 389}, 83
\item[] Rebolo, R., Zapatero Osorio, M.R. \& Martin, E.L., 1995, {\it Nature}, {\bf 377}, 83
\item[] Rogers, F. \& Young D., 1997,  {\it Phys. Rev. E} {\bf 56}, 5876
\item[] Ruiz, M.-T., Leggett, S. \& Allard, F., 1997, {\it Astroph. J.}, {\bf 491}, L107
\item[] Salpeter, E.E., 1954,  {\it Austral. J. Physics}, {\bf 7}, 373
\item[]  Saumon, D. \& Chabrier, G., 1991, {\it Phys. Rev. A} {\bf 44}, 5122; 1992, {\it Phys. Rev. A} {\bf 46}, 2084
\item[]  Saumon, D., Bergeron, P., Lunine, L.I., Hubbard, W.B., and
Burrows, A., 1994, {\it Astroph. J.}, {\bf 424}, 333
\item{} Saumon, D., Hubbard, W.B., Chabrier, G. \& Van Horn, H.M., 1992,
 {\it Astroph. J.} {\bf 391}, 827
\item{} Saumon, D., Chabrier, G., Wagner, D.J., and Xie, X., 1998, {\it Phys.
Rev.}, submitted
\item[] Schatzman, E., 1948, {\it J. Phys. Rad}, {\bf 9}, 46
\item[] Silk, J., 1977,  {\it Astroph. J.} {\bf 211}, 638
\item[] Stevenson, D., 1991, {\it ARA\&A} {\bf 29}, 163
\item[] Stevenson, D.  \& Ashcroft, N. W., 1974, {\it Phys. Rev. A} {\bf 9}, 782
\item[] Stevenson, D.J. \& Salpeter, E.E., 1977, {\it Astroph. J. Supp.} {\bf 35}, 221
\item[] Tsuji, T., Ohnaka, K., and Aoki, W., 1996, {\it Astron. \& Astroph.} {\bf 305}, L1
\item[] Weir, S.T., Mitchell, A.C., \& Nellis, W.J., 1996, {\it Phys. Rev. Lett.} {\bf 76}, 1860
\item[] Wigner, E. \& Huntington, H.B. 1935, {\it J. Chem. Phys.} {\bf 3}, 764
\end{harvard}

\end{document}